\documentclass[intlimits,twoside,a4paper]{article}

\usepackage[cp1251]{inputenc}

\usepackage[eqsecnum]{cmpj3}

\usepackage{bm}
\usepackage{color}
\usepackage{colortbl}

\issue{2023}{26}{2}{23705}
\doinumber{10.5488/CMP.26.23705}

\title[Spectral properties of a broadband far infrared photodetector with a new design of active region]%
{Spectral properties of a broadband far infrared photodetector with a new design of active region}
\author[M.~V.~Tkach, Ju.~O.~Seti, O.~M.~Voitsekhivska, V.~V.~Hutiv]{M.~V.~Tkach\orcid{0000-0003-2620-2236}\footnote{E-mail: \email{m.tkach@chnu.edu.ua}}, 
Ju.~O.~Seti\orcid{0000-0001-5576-8031}, O.~M.~Voitsekhivska\orcid{0000-0003-2118-231X}, V.~V.~Hutiv\orcid{0000-0003-2076-5222}}
\address{Yuriy Fedkovych Chernivtsi National University, 2 Kotsyubinsky Str., 58012 Chernivtsi, Ukraine}

\date{Received October 13, 2022, in final form January 07, 2023}

\begin{document}

\maketitle

\begin{abstract}
A quantum theory of spectral parameters and oscillator strengths of quantum transitions in an active
 region, which contains cascades of wide quantum wells with a complicated potential profile is developed.
 A new spatial design of the cascade is calculated and proposed with such an asymmetric arrangement
 of the wells and barriers, in which, without an applied electric bias, the magnitudes of
 oscillator strengths are considerable and one-way resonant-tunneling transport of electrons
 is observed. As a result, it becomes possible to ensure a successful functioning of the
 broadband photodetector in the far IR range.
\keywords electron, spectrum, quantum transition, photodetector, resonant-tunneling transport
%
\end{abstract}

\section{Introduction}

The research and application of isotropic nanostructures (GaAs/Al$_{x}$Ga$_{1-x}$As, In$_{x} $Ga$_{1-x}$As/Al$_{y}$In$_{1-y}$As, GaInAs/AlAsSb) as active regions of IR electromagnetic field detectors, started in the 1980 and 1990, and continues intensively until now. The physical basis of their operation are the bound states of electrons that arise in multi-quantum wells (MQW) due to the spatial quantization. Just intersubband quantum transitions of electrons, accompanied by absorption or emission of an electromagnetic field, enable  successful operation of different nanodevices: quantum well IR photodetectors (QWIPs), quantum cascade detectors (QCD) and lasers~(QCL) also operating in IR range of frequencies.

In the initial theoretical and experimental papers~\cite{Kam74,Esa77,And82,Chi83,Coo84,Coo86}, the main principles of the physics of confinement, quantum transitions, and electron tunneling through the MQWs were established. They became the background for experimental fabrication of the first photoconductive~\cite{Lev87,Ban92,Luo05,Gun05} and photovoltaic~\cite{Sch91,Sch92,Sch93,Sch97} QWIPs, produced on isotropic nanostructures. The main difference between them~\cite{Sch07} is that, due to the different designs of active regions, photoconductive QWIPs operate only in a constant electric field (bias), which promotes the extraction of electrons from the excited states of quantum wells and their one-way transport through the structure. In photovoltaic QWIPs such transport of electrons happens due to the asymmetric MQW structure even at zero bias.

Early in 2000, experimental QWIPs based on double and triple compounds GaAs/AlGaAs, \mbox{InGaAs/AlInAs}, GaInAs/AlAsSb detected an electromagnetic field from the middle- to far IR range. Those devices, which operated at medium wavelengths reached ``commercial maturity'', being used both for civil and military purposes. However, due to conduction band offset (CBO) limitations, such detectors are not capable of reaching the near IR range in the actual window  (1,3~{\textmu}m--1,55~{\textmu}m) used for fiber-optic communication. This turned out to be one of the main factors that prompted Suzuki and colleges~\cite{Suz97,Suz98} to be the first who paid attention to nitride compounds, which soon~\cite{Hof03, Bau06, Bau07, Bau09} became the basic anisotropic elements of nanodevices in the IR range (QWIP, QCD, QCL), being actively researched and used until now.

Unlike isotropic arsenide material structures with cubic unit cells, there are nitride compounds  (GaN/AlN, GaN/AlGaN, AlInN/GaInN) with anisotropic hexagonal unit cells. They possess an internal polarization electric field, having a large CBO (2~eV) and longitudinal optical phonons of high energy (90~meV). These characteristics of nitride structures made it possible to master two extreme subranges of the IR spectrum, which are fundamentally inaccessible to arsenide compounds --- near (1~{\textmu}m--3~{\textmu}m) and very far terahertz range in the region of frequencies close to optical phonons. Nitride compounds, such as components of QWIP, QCD and QCL active elements, have a significant drawback --- a large discrepancy between the sizes of elementary cells of the contacting layers creating the quantum wells and barriers. This leads to a considerable density of defects and cracks, greatly impairing the main operating parameters of nanodevices. In order to get rid of this influence, in particular, the proper design and selection of such contacting materials are used, which, having opposite directions of lattice deformation, give a rather small resulting deformation. The other approach is to use quantum dots (QDs) (for example, GaN/AlN) instead of quantum wells in active regions of superlattices~\cite{Mou03, Tch05, Var09}, which increases the sensitivity of the detectors tenfold.

A permanent alternative to QWIPs has been QCDs, which from the first years of their appearance~\cite{Gen04, Gen05, Hofs06} until now compete in terms of basic operating parameters. The advantages, disadvantages and problems associated with this type of nanodetectors are described and analyzed in detail in reviews~\cite{Gio09, Bee13, Gub22, Dur16}. Both arsenide~\cite{Gio09} and nitride~\cite{Bee13} QCD cascades mainly consist of one (two) active wells and MQWs of the extractor. In the model of electron effective mass, the QCD cascades operate as follows: an electron transits from the ground level to the first excited level in a doped active well due to the absorption of photon energy. From the excited level, due to the resonant-tunneling (with some small probability) it falls directly to the same level of the active well of the next cascade. Relaxing in the extractor with successive radiation of optical phonon (with big probability) using the steps of the ``phonon ladder'', it gets to the ground level of the active well of the next cascade.

However, now both types of nanodetectors (QWIPs and QCDs) based on arsenide and nitride have covered the whole IR range. The need to improve their operating parameters (detectivity, sensitivity) at different temperature regimes remains relevant and is constantly in the center of a research attention. Both nitride (polar, semi-polar, and nonpolar)~\cite{Gub22, Dur16, Men19} and arsenide~\cite{Hua17, Hua18, Zha19, Kim19} photodetectors are permanently modified and improved. In these and in many other papers, different materials of active elements and substrates are used instead of quantum well (QW) layers: superlattices of quantum dots or disks, nanowires, etc. Furthermore, different designs of active regions are proposed now.

Intensive development of modern nano technologies in civil and military activities (management, ecology, science, space, navigation, etc.) requires the detectors (QWIP, QCD), which successfully operate in the entire IR range. The photodetectors of the far and very far IR range must have a sufficient magnitude of the relative width $(\eta=\Delta\lambda/\lambda )$ of electromagnetic field absorption band for an effective operation in the corresponding window of atmosphere transparency. This is not a simple task for experimental implementation, especially in the range of terahertz frequencies, where photon energies are close to the energies of optical phonons of arsenide (30~meV) and nitride (90~meV) compounds. Various approaches to improve the parameters of broadband photodetectors are used now. Among them is a proper selection of materials for the wells and barriers of active regions and a carefully applied temperature regime for their growth with an accurately calculated design of the potential profile of nanostructures~\cite{Gub22, Zho19, Yan20}.

In this paper, a quantum theory of electron spectrum and oscillator strengths of quantum transitions is proposed for a new model of a far- and ultra-far IR broadband nanodetector with the design of active region that contains cascades of different QWs with a complicated potential profile. Since in the studied structure there is no ``phonon ladder'' and bias potential, the dark current is absent. Thus, the response current must significantly increase in such photodetector.

\section{Energy spectrum and wave functions of an electron in a cascade of QWs with complicated potential profile}
\label{sec_2}

In order to maximally avoid the appearence of a considerable dark current due to the electric bias applied to the active region of a broadband nanodetector operating in the far IR range, we are going to study such a design of its cascades, which does not require a bias. As far as we know, it has not been studied yet. Thus, the active region of detector consists of $N$ identical cascades, each of which, in its turn, contains one ordinary structureless wide QW and two groups of wide QWs with the same heights of potential barriers ($U$) and half widths of wide wells ($D$). As for the thicknesses ($L$) of the interwell barriers, to avoid the cumbersome analytics in the further theory and complicated numerical calculations, we consider a model in which the mutual insignificant influence of the neighboring wide QWs is neglected. It is assumed that these QWs are located at a considerably large distance ($L$). According to the experimental estimations $L\geqslant 50$ nm~\cite{Luo05}.

The applied model made it possible to calculate and investigate the properties of the spectrum of the entire cascade, as a superposition of independent spectra of three types of separate wide QWs (with and without inner barriers and wells).

\begin{figure}[!t]
\centerline{\includegraphics[width=1.0\textwidth]{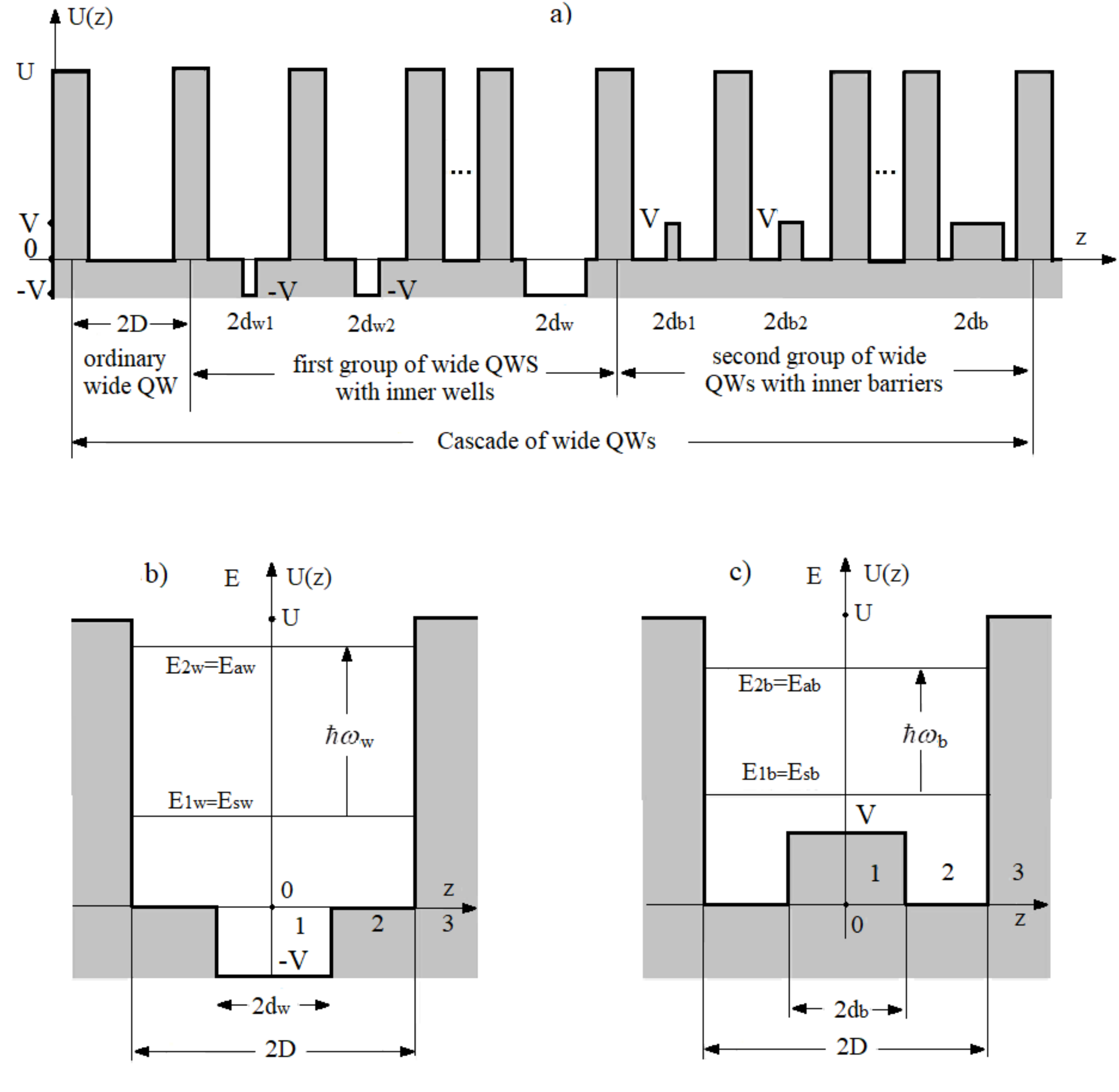}}
\caption{Potential profiles and schemes of energy levels. Separate cascade of weakly bound wide QWs with equal widths (2D), which consists of an ordinary wide QW and two groups of wide QWs (a). The first group contains only inner wells while the second one has got only inner barriers. Separate wide QW with an inner well (b); separate wide QW with an inner barrier (c).} \label{fig1}
\end{figure}

 In figure~\ref{fig1}, the potential profiles for the cascade (a), the first (b) and the second (c) groups of wide QWs are presented. The parameters $U$ and $D$ are chosen so that the weakly bound wide QWs contain only two energy levels, each. Their energies are: $E_{s}^{(\pm)}$ --- that of the ground (symmetric) states and  $E_{a}^{(\pm)}$ --- that of the first excited (asymmetric) states. The parameters $V$, $d_{n^{-}}$, $d_{n^{+}}$, are chosen to ensure the tunneling transport of excited electrons in one-way direction of the cascade (from left to right, to be specific). According to the theory developed further, their correct choice optimizes the characteristics of the electromagnetic field absorption bands.

In order to clarify under what conditions and at which parameters of nanostructure the operation of the studied broadband photodetector cascade is possible, a theory of the energy spectrum, wave functions and oscillator strengths of electron quantum transitions under the action of an electromagnetic field in the IR range is developed. Figure~\ref{fig1} shows that the potential profile contains one wide QW without an inner structure and two groups of wide QWs with an inner structure. The first group contains ($n_{w}^{-}$) inner wells with potentials  ($-V$), half widths  ($d_{n^{-}}$) and electron effective mass $m^{(-)}=m_{1w}$.  The second group contains ($n_{b}^{+}$) inner barriers with potentials ($V$), half widths ($d_{n^{+}}$) and effective mass $m^{(+)}=m_{1b}$. Simplifying the cumbersome calculations, we neglect the mutual  insignificant influence of different wide QWs, and introduce compact notations
\begin{equation} \label{Eq1}
    U^{(\pm)}(z)=\left\{
  \begin{array}{lcr}
    \pm V ;  \\
    0 ; \\
    U ;
  \end{array}\right.
  m^{(\pm)}(z)=\left\{
  \begin{array}{lcr}
    m_{1}^{(\pm)}=m_{1}^{{b \choose w}} ;  \\
    m_{2}^{(\pm)}=m_{w} ; \\
    m_{3}^{(\pm)}=m_{b} ;
  \end{array}\right.
  \begin{array}{lcr}
   0\leqslant\mid z \mid\leqslant d,  \\
   d\leqslant\mid z \mid\leqslant D, \\
   D\leqslant\mid z \mid\leqslant \infty.
  \end{array}
   \begin{array}{lcr}
   (1)  \\
   (2) \\
   (3)
  \end{array}
\end{equation}
Numerical values of the potentials  ($U$, $V$), electron effective mass in wells ($m_{w}$, $m_{1w}$) and barriers ($m_{b}$, $m_{1b}$), as well as the half width of wide QWs ($D$), small inner wells ($d_{n^{-}}$) and small barriers ($d_{n^{+}}$) will be further determined after the solution of two independent Schr\"{o}dinger equations
\begin{equation} \label{Eq2}
  \left\{-\frac{\hbar^{2}}{2}\frac{\partial}{\partial z}\frac{1}{m(z)^{(\pm)}}\frac{\partial}{\partial z}+U^{(\pm)}(z)\right\}\psi^{(\pm)}(z)=E^{(\pm)}\psi^{(\pm)}(z).
\end{equation}
Further, the spatial symmetry of three types of wide QWs is taken into account (figure~\ref{fig1}~a, b, c). In addition, for the successful operation of the detector, the physical requirement is ensured that in the subbarrier region of energies [$V \leqslant E^{(\pm)} \leqslant U$] only two energy levels are present in each QW: that with the energies of the ground symmetric $\left|s \right>$-states  [$E_{1}^{(\pm)}=E_{s}^{(\pm)}$],   and that with the energies of the excited asymmetric $\left|a \right>$-states [$E_{2}^{(\pm)}=E_{a}^{(\pm)}$]. At the same time, the energy levels of $\left|a \right>$-states should be located near the tops of each wide QW in order to ensure an easy tunneling of the electrons from this state of the previous QW into the next QW, producing an electric current in the active region. Thus, the following condition should be fulfilled
\begin{equation} \label{Eq3}
  \Delta E_{U2}=U-E_{2}^{(\pm)}\ll U.
\end{equation}
At first, we consider the equation (\ref{Eq2}) for the wide QW with an inner barrier of an arbitrary fixed half-width $d$ in the actual energy range: $V\leqslant E_{s,a}^{(+)}\leqslant U$. Then, in the region $0\leqslant z\leqslant \infty$ according to the quantum mechanics, the solution of equation (\ref{Eq2}) for the odd ground state is a symmetric $(s)$ wave function
\begin{equation} \label{Eq4}
\psi_{s}^{(+)}(0\leqslant z\leqslant \infty)=\left\{
\begin{array}{lcr}
    A_{s1}^{(+)}\cos(k_{1}z);  \\
    A_{s1}^{(+)}\cos(k_{1}d)\frac{\sin[k_{2}(D-z)]}{\sin[k_{2}(D-d)]}+A_{s3}^{(+)}\re^{-\gamma D}\frac{\sin[k_{2}(d-z)]}{\sin[k_{2}(d-D)]}; \\
    A_{s3}^{(+)}\exp({-\gamma z});
\end{array}\right.
 \begin{array}{lcr}
    0\leqslant z\leqslant d,  \\
    d\leqslant z\leqslant D, \\
    D\leqslant z\leqslant \infty,
\end{array}
\end{equation}
and for the even excited state is an asymmetric $(a)$ wave function
\begin{equation} \label{Eq5}
\psi_{a}^{(+)}(0\leqslant z\leqslant \infty)=\left\{
\begin{array}{lcr}
    A_{a1}^{(+)}\sin(k_{1}z);  \\
    A_{a1}^{(+)}\sin(k_{1}d)\frac{\sin[k_{2}(D-z)]}{\sin[k_{2}(D-d)]}+A_{a3}^{(+)}\re^{-\gamma D}\frac{\sin[k_{2}(d-z)]}{\sin[k_{2}(d-D)]}; \\
    A_{a3}^{(+)}\exp({-\gamma z});
\end{array}\right.
 \begin{array}{lcr}
    0\leqslant z\leqslant d,  \\
    d\leqslant z\leqslant D, \\
    D\leqslant z\leqslant \infty,
\end{array}
\end{equation}
where
\begin{equation} \label{Eq6}
  k_{1}=\hbar^{-1}\sqrt{2m_{b}\left[E_{s,a}^{(+)}-V\right]}; \quad
  k_{2}=\hbar^{-1}\sqrt{2m_{w}E_{s,a}^{(+)}}; \quad
  \gamma=\hbar^{-1}\sqrt{2m_{b}\left[U-E_{s,a}^{(+)}\right]}.
\end{equation}
From the continuity conditions for the wave functions (\ref{Eq4}), (\ref{Eq5}) and their densities of currents at interfaces of the structure
\begin{equation} \label{Eq7}
\left\{
  \begin{array}{lr}
  \psi_{s,a}^{(+)}(z=d-0)=\psi_{s,a}^{(+)}(d+0); \\ \\
  \psi_{s,a}^{(+)}(z=D-0)=\psi_{s,a}^{(+)}(D+0);
  \end{array}\right.
   \begin{array}{lcr}
   \frac{1}{m_{b}}\frac{\rd\psi_{s,a}^{(+)}(z)}{\rd   z}\Bigl|_{z=d-0}=\frac{1}{m_{w}}
   \frac{\rd\psi_{s,a}^{(+)}(z)}{\rd z}\Bigl|_{z=d+0}\,, \\ \\
   \frac{1}{m_{w}}\frac{\rd\psi_{s,a}^{(+)}(z)}{\rd z}\Bigl|_{z=D-0}=\frac{1}{m_{b}}
   \frac{\rd\psi_{s,a}^{(+)}(z)}{\rd z}\Bigl|_{z=D+0} \,,
   \end{array}
\end{equation}
 two independent systems of two linear homogeneous equations are found. Using them, the energies $E_{s}^{(+)}$ of the ground and all odd (symmetric) states are obtained from the first dispersion equation
\begin{equation} \label{Eq8}
  \cos^{2}(k_{2}\Delta)\left[1+\frac{\gamma m_{w}}{k_{2}m_{b}}\tan(k_{2}\Delta)\right]
  \left[1+\frac{k_{1}m_{w}}{k_{2}m_{b}}\tan(k_{1}d)\tan(k_{2}\Delta)\right]=1,
\end{equation}
and the energies  $E_{a}^{(+)}$ of all excited odd (asymmetric) states are obtained from the second dispersion equation
\begin{equation} \label{Eq9}
  \cos^{2}(k_{2}\Delta)\left[1+\frac{\gamma m_{w}}{k_{2}m_{b}}\tan(k_{2}\Delta)\right]
  \left[1+\frac{k_{1}m_{w}}{k_{2}m_{b}}\cot(k_{1}d)\tan(k_{2}\Delta)\right]=1,
\end{equation}
where
\begin{equation} \label{Eq10}
  \Delta=D-d.
\end{equation}

Continuity conditions (\ref{Eq7}) together with the normality of wave function
\begin{equation} \label{Eq11}
  2\int_{0}^{\infty}\left|\psi_{s,a}^{(+)} \right|^{2}\rd z=1
\end{equation}
determine all coefficients $A^{+}$, and, hence, the electron wave functions in wide QW with a small inner barrier. In a compact form, these functions can be presented as:
\begin{equation} \label{Eq12}
\psi_{s}^{(+)}(z)=\frac{1}{N_{s}^{(+)}}\left\{
\begin{array}{lcr}
    \cos(k_{1}z);  \\
    \frac{\cos(k_{1}d)\sin[k_{2}(D-z)]+B_{s}^{(+)}\sin[k_{2}(z-d)]}{\sin[k_{2}\Delta]}; \\
    B_{s}^{(+)}\exp[{-\gamma (z-D)}];
\end{array}\right.
 \begin{array}{lcr}
    0\leqslant z\leqslant d,  \\
    d\leqslant z\leqslant D, \\
    D\leqslant z\leqslant \infty,
\end{array}
\end{equation}
\begin{equation} \label{Eq13}
\psi_{a}^{(+)}(z)=\frac{1}{N_{a}^{(+)}}\left\{
\begin{array}{lcr}
    \sin(k_{1}z);  \\
    \frac{\sin(k_{1}d)\sin[k_{2}(D-z)]+B_{a}^{(+)}\sin[k_{2}(z-d)]}{\sin[k_{2}\Delta]}; \\
    B_{a}^{(+)}\exp[{-\gamma (z-D)}];
\end{array}\right.
 \begin{array}{lcr}
    0\leqslant z\leqslant d,  \\
    d\leqslant z\leqslant D, \\
    D\leqslant z\leqslant \infty.
\end{array}
\end{equation}
Here,
\begin{eqnarray} \label{Eq14}
 N_{s}^{(+)}&=&\sqrt{2} \left [\frac{2dk_{1}+\sin(2dk_{1})}{4k_{1}}+\frac{(B_{s}^{(+)})^{2}}{2\gamma}
    +\frac{1}{4k_{2}\sin^{2}(k_{2}\Delta)}\right.\Big\{ \left[(B_{s}^{(+)})^{2}+\cos^{2}(k_{1}d) \right] \nonumber\\
 && \times \left[2k_{2}\Delta-\sin(2k_{2}\Delta)\right]
 +4B_{s}^{(+)}\cos(k_{1}d)\left[\sin(k_{2}\Delta)-k_{2}\Delta\cos(k_{2}\Delta)\right] \Big\}\Biggl]^{1/2},
\end{eqnarray}
\begin{eqnarray} \label{Eq15}
 N_{a}^{(+)}&=&\sqrt{2} \left [\frac{2dk_{1}+\sin(2dk_{1})}{4k_{1}}+\frac{(B_{a}^{(+)})^{2}}{2\gamma}
    +\frac{1}{4k_{2}\sin^{2}(k_{2}\Delta)}\Big\{ \left[(B_{a}^{(+)})^{2}+\sin^{2}(k_{1}d) \right]\right. \nonumber\\
 && \times \left[2k_{2}\Delta-\sin(2k_{2}\Delta)\right]
 +4B_{a}^{(+)}\sin(k_{1}d)\left[\sin(k_{2}\Delta)-k_{2}\Delta\cos(k_{2}\Delta)\right] \Big\}\Biggl]^{1/2},
\end{eqnarray}
are the normality coefficients. The functions $B_{s,a}^{(+)}$ are as follows
\begin{equation} \label{Eq16}
  B_{s}^{(+)}=\cos(k_{1}d)\cos(k_{2}\Delta)-\frac{k_{1}m_{w}}{k_2m_{b}}\sin(k_{1}d)\sin(k_{2}\Delta),
\end{equation}
\begin{equation} \label{Eq17}
  B_{a}^{(+)}=\sin(k_{1}d)\cos(k_{2}\Delta)+\frac{k_{1}m_{w}}{k_2m_{b}}\cos(k_{1}d)\sin(k_{2}\Delta).
\end{equation}
Solving, in a similar way, the Schr\"{o}dinger equation (\ref{Eq2}) for the wide QW with the inner well in the energy range  $0\leqslant E_{s,a}^{(-)}\leqslant U$, two independent equations are obtained
\begin{equation} \label{Eq18}
  \cos^{2}(k_{2}\Delta)\left[\frac{\gamma m_{w}}{k_{2}m_{b}}\tan(k_{2}\Delta)+1 \right]
  \left[1-\frac{k_{1w}}{k_{2}}\tan(k_{2}\Delta)\tan(k_{1w}d) \right]=1 ,
\end{equation}
\begin{equation} \label{Eq19}
  \cos^{2}(k_{2}\Delta)\left[\frac{\gamma m_{w}}{k_{2}m_{b}}\tan(k_{2}\Delta)-1 \right]
  \left[1+\frac{k_{1w}}{k_{2}}\tan(k_{2}\Delta)\tan(k_{1w}d) \right]=1,
\end{equation}
for the energies $E_{s}^{(-)}$ and $E_{a}^{(-)}$ in symmetric and asymmetric states, respectively. Here, $k_{1w}=\hbar^{-1}\times \sqrt{2m_{w}[E^{(-)}+V]}$, and the values $k_{2}$ and $\gamma$ are the same as in formulae (\ref{Eq6}). Applying the limit $V\rightarrow 0$, $d\rightarrow0$ in formulae~(\ref{Eq18}) and~(\ref{Eq19}), two known equations are obtained for the energies of the ground and first excited states of an ordinary wide QW
\begin{equation} \label{Eq20}
  \cos^{2}(k_{2}D) \left[\frac{\gamma m_{w}}{k_{2}m_{b}}\tan(k_{2}D)\pm 1 \right]=1; \qquad\qquad
  \begin{cases}
    + \text{\space for $s$-state},\\
    - \text{\space for $a$-state}.
\end{cases}
\end{equation}
In analogy with the analytical calculations of the wave functions presented above, for the wide QW with inner barrier, the exact expressions are obtained for the wave functions of the electron in the ordinary wide QW and in QW with inner well. They are not presented here explicitly to avoid cumbersome analytical expressions.

\section{Oscillator strengths of quantum transitions. Optimization of the parameters of QW cascades in broadband photodetector operating in the far IR range}

For a successful operation of a broadband photodetector in the far IR range, each cascade must have a spatial design of three types of wide QWs, which ensures that the positions of the energy levels and oscillator strengths of quantum transitions satisfy the conditions (\ref{Eq3}), (\ref{Eq21}), (\ref{Eq22}). It was already noted that for a sufficient tunneling of electrons between the wide QWs, the condition (\ref{Eq3}) must be fulfilled, i.e., the highest energy level of the bound excited state of wide QW should be located very close to its top. Secondly, the tunneling transport of electrons in the cascade must occur only in one-way direction, since
\begin{equation} \label{Eq21}
  \Delta E_{2,n^{(\pm)}}=E_{2,n^{(\pm)}}-E_{2,n^{(\pm)}+1}\ll E_{2,n^{(\pm)}}, \qquad [n^{(\pm)}=1,2,3,\dots] .
\end{equation}
It means that the difference between the energies of the excited states of each previous and next QW should be positive and not large. Finally, as the absorption coefficient of the electromagnetic field should be sufficiently big in the required energy range and it is proportional to the oscillator strength of quantum transition $(f_{sa})$, then a condition
\begin{equation} \label{Eq22}
  f_{sa}\geqslant0.5
\end{equation}
must be fulfilled in the whole range of the detected energy.

The theory developed above makes it possible to determine the energy levels and wave functions, which are necessary for the calculation and optimization of the spatial design of wide QWs. The analytical expression for the oscillator strength of quantum transition between the states $ \left|s \right>$ and $\left|a\right>$ with the energies $E_{s}$ and $E_{a}$ is known
\begin{equation} \label{Eq23}
  f_{sa}=\frac{2m_{0}}{\hbar^{2}}(E_{a}-E_{s})\left|\langle\psi_{s}(z)|z|\psi_{a}(z)\rangle\right|^{2},
\end{equation}
but it is valid only for a free electron with a spatially constant mass $m_{0}$. Therefore, for the cascade under study, when mass $m(z)$ is position-dependent, formula (\ref{Eq23}) should be generalized in such a way that the condition of orthonormality of wave functions  [$\psi_{s}^{\pm}(z),\psi_{a}^{\pm}(z)$] is not violated. These functions were obtained in the section~\ref{sec_2} as the exact solutions of the Schr\"{o}dinger equations (\ref{Eq2}) with the Hermitian Hamiltonian of the electron with different effective masses in different layers of the respective wide QWs with inner barriers or wells.

From physical considerations, it is clear that, satisfying this requirement, such generalization can be carried out if, instead of $m_{0}$ effective mass,  we introduce the one which is correlated over the separate wide QW
\begin{equation} \label{Eq24}
 m^{(\pm)}=m_{0}\mu^{(\pm)},
\end{equation}
where $\mu^{(\pm)}$ is the dimensionless correlated mass, which takes into account the probability of electron location [$P_{\ell=1,2,3}^{(\pm)}$] in each of the three regions of the respective wide QW, as well as the symmetry in both states between which the quantum transition occurs
\begin{equation} \label{Eq25}
 \mu^{(\pm)}=P_{1}^{(\pm)}m_{1}^{(\pm)}+P_{2}^{(\pm)}m_{2}^{(\pm)}+P_{3}^{(\pm)}m_{3}^{(\pm)}.
\end{equation}
Here, $m_{\ell}^{(\pm)}$ is a dimensionless (in $m_{0}$ units) effective mass in $\ell$-th medium with internal barriers ($+$) or a well ($-$), and $P_{\ell}^{(\pm)}$ is averaged over both states probability of electron location in $\ell$-th medium of wide QW
\begin{eqnarray} \label{Eq26}
 &&P_{1}^{(\pm)}=\int_{0}^{d}\left[\left|\psi_{1s}^{(\pm)}(z) \right|^{2}+\left|\psi_{1a}^{(\pm)}(z) \right|^{2} \right]\rd z,\qquad P_{2}^{(\pm)}=\int_{d}^{D}\left[\left|\psi_{2s}^{(\pm)}(z) \right|^{2}+\left|\psi_{2a}^{(\pm)}(z) \right|^{2} \right]\rd z, \nonumber\\
 && P_{3}^{(\pm)}=\int_{D}^{\infty}\left[\left|\psi_{3s}^{(\pm)}(z) \right|^{2}+\left|\psi_{3a}^{(\pm)}(z) \right|^{2} \right]\rd z.
\end{eqnarray}
Finally, the oscillator strength of quantum transition $(f_{sa}^{\pm})$ for the electron with position-dependent effective mass in wide QW is given by an analytical expression, which is further used for numeric calculations
\begin{equation} \label{Eq27}
 f_{sa}^{\pm}=\frac{E_{a}^{(\pm)}-E_{s}^{(\pm)}}{\mathrm{Ry}} \ \frac{\mu^{(\pm)}}{a_{B}^{2}}
 \left| \langle\psi_{s}^{(\pm)}(z)|z|\psi_{a}^{(\pm)}(z)\rangle \right|^{2},
\end{equation}
where Rydberg energy is introduced for the sake of convenience  $(\mathrm{Ry}=\hbar^{2}/2m_{0}a_{\rm B}^{2})$ and $a_{\rm B}$ is the Bohr radius.

Using the developed theory, the spectral parameters of wide QW cascades were calculated and optimized depending on the magnitudes of the potentials $U$ and $V$, and sizes $D$ and $d$.  GaAs/Al$_{x}$Ga$_{1-x}$As wide QWs with small inner barriers Al$_{y}$Ga$_{1-y}$As and small inner wells Ga$_{y}$In$_{1-y}$As are the components of the cascade. Such a  choice of compounds is caused by the reason that, as is known, they have close sizes of elementary cells, which protects the cascade from the formation of unwanted stresses. The simple linear relationships between the concentrations ($x$, $y$) of the elements, electron potential energies ($U$, $V$) and its effective mass ($m$) in all three regions of wide QW are as follows:
\begin{eqnarray} \label{Eq28}
 &&U=1.1x_{\rm Al}\,(\rm eV), \quad V_{b}=1.1x_{\rm Al}\,(\rm eV); \quad |V_{w}|=0.85y_{\rm Ga}\,(\rm eV);\nonumber\\
 &&m_{\rm GaAs}^{w}=0.067;\quad m_{{\rm Al}_{x}{\rm Ga}_{1-x}{\rm As}}^{b}=0.067+0.083x_{\rm Al}; \quad m_{{\rm Al}_{y}{\rm Ga}_{1-y}{\rm As}}^{w}=0.022+0.045y_{\rm Ga}.
\end{eqnarray}
For a successful operation of broadband photodetector it is necessary to create a specific arrangement of wide QWs with such half widths $(d)$ of inner barriers and wells, which would ensure the conditions~(\ref{Eq3}),~(\ref{Eq21}),~(\ref{Eq22}) with respect to spectral parameters [$E_{a}^{(\pm)}, \hbar\omega^{(\pm)}$] and oscillator forces $(f_{sa}^{\pm})$ at fixed values of $U$, $V$, $D$. An example of the calculated values [$E_{a}^{(\pm)}$, $\hbar\omega^{(\pm)}$, $f_{sa}^{(\pm)}$] for the fifteen wide QWs (with different $d$), being the elements of a separate photodetector cascade, are presented in table~\ref{tbl1}. The magnitudes of potentials $U$, $V_{b}$, $V_{w}$ were calculated using the relationships (\ref{Eq28}) at the values of concentrations ($x$, $y$) indicated in the caption to table~\ref{tbl1}. The respective effective masses in all regions of the wide QW were defined at the same concentrations. This proves that the width of the range of some peak energies of the electromagnetic field absorbed by the cascade of sequentially located wide QWs of both groups with the same $U$,  $V$, $D$, is limited by the sizes of inner wells or barriers $d_{n^{(\pm)}}<D/2$. This is caused by violation of the conditions (\ref{Eq21}) and (\ref{Eq22}) if $d_{n^{(\pm)}}$ tends from below to $D/2$ for wide QWs with inner wells and inner barriers, respectively. In table~\ref{tbl1}, the parameters of wide QWs which do not satisfy the conditions~(\ref{Eq21}), (\ref{Eq22}) and, therefore, are not included in the cascade structure are highlighted. Thus, the operating cascade of the studied example should contain only ten wide QWs arranged as shown in figure~\ref{fig1}. Initially, there are five wide QWs (with half widths of inner wells varying in the interval $0 \ {\rm nm}\leqslant d^{-}\leqslant 1 \ {\rm nm}$), followed by five wide QWs (with half widths of the inner barriers in the interval $0.25 \ {\rm nm}\leqslant d^{+}\leqslant 1.25 \ {\rm nm}$). We should note that, as the calculations and analysis prove, if the half widths ($d^{\pm}$) of the inner wells and barriers increase and exceed the value $D/2$, the energy [$E^{(\pm)}_{a}$] of the corresponding excited states only increase. As a result, a reverse electron tunneling may occur in the structure. Thus, such wide QWs should not be included into the cascade.

\begin{table}[htb]
\caption{Spectral parameters [$E_{a,s}^{(\pm)}, \hbar\omega^{(\pm)}$] and oscillator strengths of quantum transitions ($f_{sa}^{\pm}$) as functions of half widths [$d^{(\pm)}$] of inner
wells and barriers at $D=3.5$~nm and magnitudes of potentials, which are determined by the respective concentrations of elements $U=160$~meV ($x_{\rm Al}=0.145$), $V_{b}=40$~meV  ($x_{\rm Al}=0.0363$), $|V_{w}|=40$~meV ($y_{\rm Ga}=0.047$).}
\label{tbl1}
\vspace{2ex}
\begin{center}
\renewcommand{\arraystretch}{0}
\begin{tabular}{||c||c|c|c|c|c|c|c|c||}
\hline\hline{$d^{(\pm)}$, nm}
      & 0 & 0.25 & 0.5 & 0.75 & 1 & 1.25 & 1.5 & 1.75 \strut\\
\hline
\rule{0pt}{2pt}&&&&&&&&\\
\hline{$E_{a}^{-}$, (meV)}
      & 150.25& 150.24& 150.17& 149.79&  149.64& \cellcolor{gray} 148.99& \cellcolor{gray} 148.26&  \cellcolor{gray} 146.9\strut\\
\hline{$E_{s}^{-}$, (meV)}
      & 44.24& 40.73&  37.37& 33.62& 30.56& \cellcolor{gray} 26.89& \cellcolor{gray} 24.32& \cellcolor{gray} 20.96\strut\\
\hline{$\hbar\omega^{-}$, (meV)}
      & 106.01& 109.51& 112.8& 116.35& 119.08& \cellcolor{gray} 122.1& \cellcolor{gray} 123.94& \cellcolor{gray} 125.94\strut\\
\hline{$f_{sa}^{-}$}
      & 0.739& 0.696& 0.639& 0.58& 0.52& \cellcolor{gray} \textbf{0.476}& \cellcolor{gray} \textbf{0.442}& \cellcolor{gray} \textbf{0.434}\strut\\
\hline\hline{$E_{a}^{+}$, (meV)}
      & $-$& 148.88& 147.62& 146.63&  146.01&  145.82& \cellcolor{gray} \textbf{146.07}&  \cellcolor{gray} \textbf{146.75}\strut\\
\hline{$E_{s}^{+}$, (meV)}
      & $-$& 47.62&  50.88& 54.02& 57.06& 59.98& \cellcolor{gray} 62.77& \cellcolor{gray} 65.41\strut\\
\hline{$\hbar\omega^{+}$, (meV)}
      & $-$& 101.26& 96.74& 92.61& 88.95& 85.84& \cellcolor{gray} 83.3& \cellcolor{gray} 81.34\strut\\
\hline{$f_{sa}^{+}$}
      & $-$& 0.742& 0.728& 0.698& 0.658& 0.616& \cellcolor{gray} 0.579& \cellcolor{gray} 0.552\strut\\
\hline\hline
\end{tabular}
\renewcommand{\arraystretch}{1}
\end{center}
\end{table}

\begin{figure}[!t]
\centerline{\includegraphics[width=0.8\textwidth]{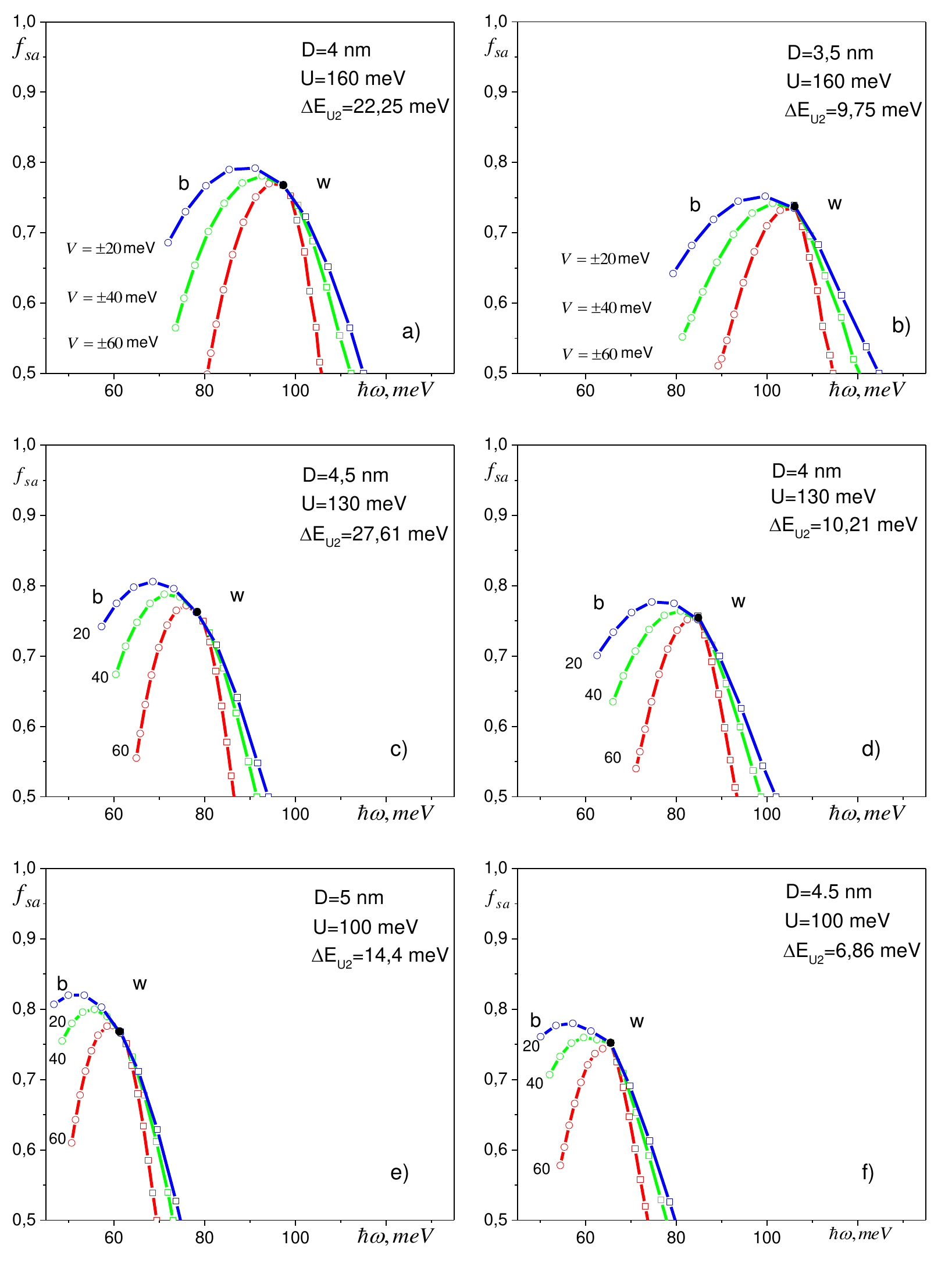}}
\caption{(Colour online) Oscillator strengths ($f_{sa}$) as functions of electromagnetic field energy ($\hbar\omega$) at different potentials ($U$, $V$) and geometric parameters ($D$, $d^{-}$, $d^{+}$) of several optimal cascades, which are presented in the panels and in table~\ref{tbl1}.} \label{fig2}
\end{figure}

Examples of the selection of potential energies ($U$, $V$) and geometric parameters ($D$, $d$) of wide QW groups, which are included into the optimized cascades of broadband photodetectors operating in three different intervals of far IR range of frequencies, are shown in figure~\ref{fig2} and in table~\ref{tbl2}. In the panels of figure~\ref{fig2} the results of the calculated oscillator forces of quantum transitions $f_{sa}$ are presented as functions of the field energy $\hbar\omega$ at three values of potential  ($U=160$~meV, 130~meV, 100~meV) which basically determine the frequency intervals ($\omega$)  or wavelengths~($\lambda$) detected by the corresponding cascades. For better visualization, $f_{sa}$ magnitudes calculated at $V={\rm const}$ are indicated by dots connected by lines. Herein, the dark dots correspond to the oscillator strengths produced by a structureless wide QW while the light dots (b) and (w) correspond to the oscillator strengths produced by the groups of wide QWs with inner barriers and wells. The functions $f_{sa}(\hbar\omega)$ are also shown at the panels of this figure at three magnitudes ($U=160$~meV, 130~meV, 100~meV) and three values of the potential energies ($V=20$~meV, 40~meV, 60~meV) of inner wells and barriers, as well as at two sizes of the half-widths $D$ of the respective QW. This illustrates which of the two cascades with the same potentials $U$ and $V$ but different sizes $D$ is more optimal due to the correct choice of the best parameters for the operating photodetector. For the clarity of analysis, numerical values $\Delta E_{U2}=U-E_{a}(V=0)$ are indicated for all six cascades.

Table~\ref{tbl2} shows the numerical values, which are necessary for further analysis. They directly characterize the optical parameters of the electromagnetic field absorbed by nine pairs of wide QW cascades (three values of $V$ at three values of $U$). They are: an average length $\lambda=(\lambda_{\rm min}+\lambda_{\rm max})/2$, width of cascade absorption band  $\Delta\lambda=\lambda_{\rm max}-\lambda_{\rm min}$ and their ratio $(\eta\%)$ as well as the average value of oscillator strengths $\overline{f}_{sa}=(\max\ell)^{-1}\sum_{\ell=1}^{\max\ell}
[f_{sa}^{(+)}(d_{\ell}^{(+)})+f_{sa}^{(-)}(d_{\ell}^{(-)})]$ of this wide QW cascade. The widths of the bands of the excited energy levels $\Delta E_{2}=E_{2}^{(-)}(d=0)-E_{2}^{(+)}(d_{\rm max})$ are also presented because together with $\Delta E_{U2}$, as it is clear from table~\ref{tbl2}, they slightly increase the magnitudes of the potential barriers above the excited levels. As a result, the tunneling inside and between the detector cascades deteriorates.

Figure~\ref{fig2} proves that at all $U$ and $V_{b}$, $V_{w}$ values, the low-frequency part of the band absorbed by an arbitrary cascade is formed by a group of wide QW, with inner barriers while the high-frequency part is formed by a structureless wide QW and group of wide QWs with inner wells. At the same time, decreasing the value of the potential $U$ at all fixed $V_{b}=|V_{w}|$ and $D$ values significantly shifts the absorption bands only to the region of lower energies, that is, to the region of longer wavelengths (table~\ref{tbl2}). From figure~\ref{fig2} and especially from table~\ref{tbl2}, it is clear that increasing $V$ at different fixed $U$ and $D$ does not significantly affect the position of the band in the spectrum, although essentially increases its width $(\Delta\lambda)$ and slightly increases the average value of oscillator strengths $f_{sa}$, that is, it improves the operating parameters of the detector.

\begin{table}[!h]
	\caption{Optical characteristics ($\lambda$, $\Delta\lambda$, $\eta$\%) and oscillator strengths ($f_{sa}$) as functions of energy and geometrical parameters of optimal cascades of wide QWs, being active elements of broadband photodetectors operating in the far IR range.}
	\label{tbl2}
	\vspace{2ex}
	\begin{center}
		\renewcommand{\arraystretch}{0}
		\begin{tabular}{||c||c|c|c|c||c|c|c||c||}
			\hline\hline
			& $V$ & $\hbar\omega_{\rm min}$ & $\hbar\omega_{\rm max}$ & $\Delta E_{2}$ & $\lambda $ & $\Delta\lambda $ & $\eta $ ,& $\overline{f}_{sa}$ \strut\\
			& (meV) & (meV) & (meV) & (meV) & (\textmu m) & (\textmu m) & \%  & $ $ \strut\\
			\hline\hline
			\rule{0pt}{2pt}&&&&&&&&\\
			
			$U=160$ meV  & 20& 80.55& 105.30& 5.94& 13.60& 3.63& 26.69& 0.639\strut\\
			$\Delta E_{2U}=$ &&&&&&&& \\ \cline{2-9}
			$=22.25$ meV & 40& 77.83& 112.17& 4.83& 13.84& 4.24& 30.64& 0.684\strut\\
			$D=4$ nm &&&&&&&& \\ \cline{2-9}
			$\overline{d_{w}=1.0}$ nm & 60& 75.76& 116.41& 4.16& 13.53& 5.72& 42.28& 0.698\strut\\
			$d_{b}=1.0$ nm &&&&&&&& \\ \cline{2-9}
			\hline\hline
			$\cellcolor{gray} U=160$ meV  & 20& 89.94& 114.03& 4.95& 12.41& 3.03& 24.42& 0.607\strut\\
			$\cellcolor{gray} \Delta E_{2U}=$ &&&&&&&& \\ \cline{2-9}
			$\cellcolor{gray} =9.75$ meV & 40& 85.84& 119.08& 4.43& 12.45& 4.04& 32.45& 0.662\strut\\
			$\cellcolor{gray} D=3.5$ nm &&&&&&&& \\ \cline{2-9}
			$\cellcolor{gray} \overline{d_{w}=0.75}$ nm & \cellcolor{gray} 60& \cellcolor{gray} 83.41& \cellcolor{gray} 121.89& \cellcolor{gray} 3.82& \cellcolor{gray} 12.52& \cellcolor{gray} 4.70& \cellcolor{gray} 37.54& \cellcolor{gray} 0.684\strut\\
			$\cellcolor{gray} d_{b}=1.0$ nm &\cellcolor{gray} &\cellcolor{gray} &\cellcolor{gray}&\cellcolor{gray}&\cellcolor{gray}&\cellcolor{gray}&\cellcolor{gray}& \cellcolor{gray} \\ \cline{2-9}
			\hline\hline
			$U=130$ meV  & 20& 64.93& 85.86& 3.32& 16.8& 4.64& 27.62& 0.678\strut\\
			$\Delta E_{2U}=$ &&&&&&&& \\ \cline{2-9}
			$=27.61$ meV & 40& 60.39& 89.62& 2.52& 17.22& 6.71& 38.97& 0.708\strut\\
			$D=4.5$ nm &&&&&&&& \\ \cline{2-9}
			$\overline{d_{w}=0.75}$ nm & 60& 57.26& 91.66& 2.04& 17.62& 8.04& 46.20& 0.732\strut\\
			$d_{b}=1.25$ nm &&&&&&&& \\ \cline{2-9}
			\hline\hline
			$\cellcolor{gray} U=130$ meV  & 20& 71.12& 92.98& 2.77& 15.41& 4.10& 26.61& 0.646\strut\\
			$\cellcolor{gray} \Delta E_{2U}=$ &&&&&&&& \\ \cline{2-9}
			$\cellcolor{gray} =10.21$ meV & 40& 68.32& 96.94& 2.46& 15.5& 5.27& 34.0& 0.691\strut\\
			$\cellcolor{gray} D=4.0$ nm &&&&&&&& \\ \cline{2-9}
			$\cellcolor{gray} \overline{d_{w}=0.75}$ nm & \cellcolor{gray} 60& \cellcolor{gray} 66.11& \cellcolor{gray} 99.09& \cellcolor{gray} 2.10& \cellcolor{gray} 15.61&\cellcolor{gray} 6.25& \cellcolor{gray} 40.04& \cellcolor{gray} 0.709\strut\\
			$\cellcolor{gray} d_{b}=1.0$ nm &\cellcolor{gray} &\cellcolor{gray} &\cellcolor{gray}&\cellcolor{gray}&\cellcolor{gray}&\cellcolor{gray}&\cellcolor{gray}& \cellcolor{gray} \\ \cline{2-9}
			\hline\hline
			$U=100$ meV  & 20& 52.52& 68.55& 1.77& 20.89& 5.53& 26.47& 0.706\strut\\
			$\Delta E_{2U}=$ &&&&&&&& \\ \cline{2-9}
			$=14.4$ meV & 40& 50.70& 71.88& 1.33& 20.89& 6.22& 29.78& 0.722\strut\\
			$D=5.0$ nm &&&&&&&& \\ \cline{2-9}
			$\overline{d_{w}=0.75}$ nm & 60& 46.75& 73.69& 1.13& 21.68& 9.65& 44.51& 0.736\strut\\
			$d_{b}=1.0$ nm &&&&&&&& \\ \cline{2-9}
			\hline\hline
			$\cellcolor{gray} U=100$ meV  & 20& 54.43& 72.07& 1.36& 20.03& 5.59& 27.91& 0.668\strut\\
			$\cellcolor{gray} \Delta E_{2U}=$ &&&&&&&& \\ \cline{2-9}
			$\cellcolor{gray} =6.86$ meV & 40& 52.07& 76.62& 1.21& 20.03& 7.64& 38.14& 0.710\strut\\
			$\cellcolor{gray} D=4.5$ nm &&&&&&&& \\ \cline{2-9}
			$\cellcolor{gray} \overline{d_{w}=0.75}$ nm & \cellcolor{gray} 60& \cellcolor{gray} 50.08& \cellcolor{gray} 78.21& \cellcolor{gray} 1.03& \cellcolor{gray} 20.31& \cellcolor{gray} 8.98& \cellcolor{gray} 44.21& \cellcolor{gray} 0.721\strut\\
			$\cellcolor{gray} d_{b}=1.0$ nm &\cellcolor{gray} &\cellcolor{gray} &\cellcolor{gray}&\cellcolor{gray}&\cellcolor{gray}&\cellcolor{gray}&\cellcolor{gray}& \cellcolor{gray} \\ \cline{2-9}
			\hline\hline
		\end{tabular}
		\renewcommand{\arraystretch}{1}
	\end{center}
\end{table}

It is important to note that at fixed values of the potentials $U$ and $V_{b}$, $V_{w}$, the intervals of half widths~$D$ in which the detector cascades operate optimally, are quite narrow. After all, the calculations show that an increasing  $D$ beyond the upper limit of operating interval does lead to a positive growth of $\overline{f}_{sa}$. At the same time, $\Delta E_{U2}$, increases sharply, which impairs or practically stops the tunneling of electrons, and thus, the functioning of the detector. Decreasing of the half width outside the operating interval has the opposite effect, i.e., it positively decreases~$\Delta E_{U2}$. However, it also reduces the average oscillator strength  $\overline{f}_{sa}$, the smaller number of quantum transitions can also stop the detector. This is clearly seen from table~\ref{tbl3}, where, for example, the calculated parameters~$\overline{f}_{sa}$ and $\Delta E_{U2}$ for the cascades of wide QWs with $U=160 $~meV,  $V_{b}=|V_{w}|=60$~meV are presented at four fixed values of half widths~($D$).

\begin{table}[!h]
\caption{Parameters of wide QWs cascade as function of $D$ at $U=160$ meV, $V=60$ meV.}
\label{tbl3}
\vspace{2ex}
\begin{center}
\renewcommand{\arraystretch}{0}
\begin{tabular}{||c||c|c|c|c||}
\hline\hline{$D$ (nm)}
      & 3.0 & 3.5 & 4.0 & 4.5   \strut\\
\rule{0pt}{2pt}&&&&\\
\hline{$\Delta E_{sa}$ (meV)}
      & 0.1& 9.75& 22.25& 40.72    \strut\\
\hline{$\bar{f}_{sa}$}
      & 0.35& 0.684& 0.698& 0.72    \strut\\
\hline\hline
\end{tabular}
\renewcommand{\arraystretch}{1}
\end{center}
\end{table}

Similar dependences are also present for the cascades with $U=130$~meV and $100$~meV. The analysis of figure~\ref{fig2} and table~\ref{tbl2} shows that among all the investigated cascades of broadband photodetectors, there are those three, whose parameters are highlighted by shading in table~\ref{tbl2}. Together, they cover a wide range of wavelengths relevant for scientific and applied research ($\lambda\approx10$~\textmu m $\div\,25$ \textmu m) of the far IR range, possessing large ratios of widths to wavelengths, which are important for these devices \mbox{($\eta\approx37.54$\% -- $44.21$\%).}

\section{Main results and conclusions}

Quantum theory of spectral parameters of broadband photodetectors functioning in the far IR range ($10$~\textmu m~$\div\,25$~\textmu m) has been developed. These devices can successfully operate, being based on the cascades of QWs with complicated potential profile without applying a constant external electric field (bias).

A new spatial design of wide QWs in the cascades is proposed. It provides a one-way tunneling of electrons through all cascades of a detector. For this purpose, the wide QWs are arranged as follows: first, an ordinary wide QW, then a group of wide QWs with shallow wells, the widths of which are successively increasing, and, then, a group of wide QWs with small barriers, with successively increasing widths, too. In such a cascade, the magnitudes of the first excited levels in each next wide QW with bigger width of inner wells or barriers slowly decrease while  the magnitude of the ground levels quickly increases. As a result, the cascade of photodetector's wide QWs covers a significant interval of the far IR range of the electromagnetic spectrum scale.

Another advantage of the proposed design of the cascade is that its QWs and barriers can be isotropic materials with close sizes of elementary cells. Unlike anisotropic materials, the isotropic structure saves the constituent elements from the formation of dislocations. This significantly impairs the operation of detectors.

Using the example of cascades with isotropic components of QWs (GaAsAs, Ga$_{y}$In$_{1-y}$As) and barriers (Al$_{x}$Ga$_{1-x}$As, Al$_{y}$Ga$_{1-y}$As), it is shown that the proper selection of concentrations ($x$, $y$), and, thus, the values of the potentials ($U$, $V_{b}$, $V_{w}$) and sizes ($D$, $d$) provides both significant oscillator strengths of quantum transitions in wide intervals of wavelengths and sufficient electron tunneling transport over all cascades. Thus, the proposed design of wide QW cascades with a complicated potential profile allows one to ensure a successful operation of broadband photodetectors in the far IR range.

\ukrainianpart

\title{Спектральні властивості з новим дизайном активної області широкосмугового нанофотодетектора далекого ІЧ-діапазону}
\author{ М. В. Ткач,  Ю. О. Сеті,   О. М. Войцехівська, В. В. Гутів}
\address{Чернівецький національний університет ім. Ю. Федьковича, вул. Коцюбинського 2, 58012 Чернівці, Україна}

\makeukrtitle

\begin{abstract}
\tolerance=3000%
Розвинена квантова теорія спектральних параметрів і сил осциляторів квантових переходів у активній
області, яка містить каскади широких квантових ям із складним потенціальним рельєфом.
Запропонований новий просторовий дизайн каскада з таким розрахованим асиметричним розташуванням
шарів ям і бар'єрів, при якому без прикладеного зміщення виникають значні величини сил осциляторів і здійс\-нюється односторонній резонансно-тунельний перенос електронів. Це дозволяє забезпечити успішне функціонування широкосмугового фотодетектора в далекому ІЧ-діапазоні.

\keywords електрон, спектр, квантовий перехід, фотодетектор, резонансно-тунельний транспорт

\end{abstract}

\end{document}